\documentstyle[epsf]{elsart}
\def\la{\mathrel{\mathpalette\fun <}}
\def\ga{\mathrel{\mathpalette\fun >}}
\def\fun#1#2{\lower3.6pt\vbox{\baselineskip0pt\lineskip.9pt
  \ialign{$\mathsurround=0pt#1\hfil##\hfil$\crcr#2\crcr\sim\crcr}}}
\begin{document}
\begin{frontmatter}
\title{
\vskip1.55cm
Reconstruction of source and cosmic magnetic field
characteristics from clusters of ultra-high energy cosmic rays}
\author{G\"unter Sigl$^1$ and Martin Lemoine$^{2,1}$}
\address{(1) Department of Astronomy \& Astrophysics\\
Enrico Fermi Institute, The University of Chicago, Chicago, IL
60637-1433\\
         (2) DARC, UPR-176, CNRS, \\
             Observatoire de Paris, 92195 Meudon C\'edex, France}
\begin{abstract}
We present a detailed Monte Carlo study coupled to a
likelihood analysis of the potential of next generation
ultra-high energy cosmic ray experiments to reconstruct
properties of the sources and the extra-galactic magnetic field.
Such characteristics are encoded in the distributions of arrival
time, direction, and energy of clusters of charged cosmic rays
above a few $10^{19}\,$eV. The parameters we
consider for reconstruction are the emission timescale, total
fluence (or power), injection spectrum, and distance of the
source, as well as the r.m.s. field strength, power spectrum, and
coherence length of the magnetic field. We discuss five generic
situations which can be identified relatively easily and allow a
reasonable reconstruction of at least part of these
parameters. Our numerical code is set up such that it can easily
be applied to the data from future experiments.

PACS numbers: 98.70.Sa, 98.62.En

Keywords: Ultra-high energy cosmic rays, cosmic magnetic fields

\end{abstract}
\end{frontmatter}


\section{Introduction}

The origin and the nature of ultra-high energy cosmic rays
(UHE CRs), with energies $E\ga10\,$EeV
($1\,{\rm EeV}=10^{18}\,$eV), are still unknown despite several 
generations of experiments, most notably the Haverah
Park~\cite{Haverah}, the Akeno Giant Air Shower
Array (AGASA)~\cite{Yoshida1,Hayashida1}, and the Fly's
Eye~\cite{Bird1} experiments. Data from this latter
seem to indicate that the UHE CR component is mainly 
composed of protons\cite{Bird1}. At these energies, protons 
cannot be confined within the Galactic magnetic field. Thus, the
isotropy of the arrival directions of most of the observed UHE
CRs~\cite{Hayashida1}, or at least the absence of a significant
correlation with the Galactic plane~\cite{Stanev},
suggests that UHE CRs are  extra-galactic in origin.

  However, such protons would leave a distinct signature in the 
energy spectrum, the so-called Greisen-Zatsepin-Kuzmin
high energy cut-off~\cite{GZK} (hereafter GZK cut-off) around
$E\simeq70\,$EeV, due to pion 
production on the cosmic microwave background by nucleons with 
$E\ga70\,$EeV. There is no strong experimental evidence for this
cut-off and the detection of particles with energies as 
high as $E\sim300\,$EeV cannot be easily explained in this
frame. No compeling astrophysical candidate for the
source of the highest energy events could be found
within $\simeq100\,$Mpc \cite{ES,HVSV},
although photopion production limits the range of 
nucleons with $E\simeq100\,$EeV to about 30 Mpc. Heavy nuclei 
would be disintegrated over similar or slightly larger
distances~\cite{Stecker}, and similar 
problems arise for the less likely option of a $\gamma$--ray 
\cite{HVSV}, for which the effective attenuation length in
electromagnetic cascades lies between 1 and 20 Mpc, depending on
the poorly known strength of the universal radio
background. Finally, neutrino primaries in general imply too 
large a flux because of their small interaction probability in 
the atmosphere \cite{SL}.

As to the theoretical models of the origin of UHE CRs, the most
conventional scenario involves 
first-order Fermi acceleration of protons in powerful 
astrophysical shocks, for instance in the hot spots of 
radio-galaxies \cite{RB}. More recently, it was suggested 
that protons could be accelerated up to $E\sim10^{21}\,$eV
in fireball models of cosmological $\gamma-$ray bursts 
\cite{Waxman1,Waxman2,Vietri1,MU}. In order to reconcile the 
observed rates of UHE CRs and cosmological $\gamma-$ray bursts 
within $D\sim30\,$Mpc, however, the arrival time of UHE CRs
would have to be spread over $\Delta\tau\ga50\,$yrs, for
instance through deflection in large-scale magnetic
fields~\cite{Waxman1,Waxman2,Waxman3}. As another class of 
models, topological defects, possible relics of
early Universe phase transitions, could release supermassive
``X'' particles with mass around the Grand Unification Scale, 
through physical processes such as collapse or annihilation
\cite{BHS}. These X particles would subsequently decay to jets
of UHE CRs, with a likely dominance of $\gamma-$rays above
$\simeq50\,$EeV, and an energy spectrum significantly harder 
than in the case of shock acceleration~\cite{Hill,Sigl}.

Charged UHE CRs, such as protons or electromagnetic cascades 
initiated by a $\gamma-$ray primary, are subject to 
energy-dependent deflection, and hence energy-dependent time 
delay, in large-scale magnetic fields. The r.m.s. strength 
$B_{\rm rms}$ and the coherence length $l_{\rm c}$ of
extra-galactic magnetic fields are thoroughly unknown, 
although they are bound by Faraday rotation data to
$B_{\rm rms}l_{\rm c}^{1/2}\la10^{-9}\,{\rm G}\,{\rm
Mpc}^{1/2}$~\cite{Kronberg}. Different authors have proposed to
use UHE CRs
to probe extra-galactic magnetic fields; some rely on 
the magnitude of the time delay and the deflection
\cite{Plaga,WC}, or on some features of the angle-time-energy 
images of UHE CRs \cite{WM,LSOS}, or even on
synchrotron loss signatures in the energy spectrum of 
electromagnetic cascades \cite{LOS}. In a previous 
paper, we discussed how information on both the extra-galactic 
magnetic field and the origin of UHE CRs could be left in 
angle-time-energy images of clusters of proton UHE CRs
\cite{LSOS}. We applied this study to a maximum likelihood
analysis of the three pairs of UHE CRs \cite{SLO}, 
that were reported by the AGASA experiment \cite{Hayashida1}. In 
order to do so, we devised a Monte-Carlo code that follows the 
propagation of UHE protons and calculates a likelihood as a 
function of the parameters characterizing the origin of these 
UHE CRs and the intervening magnetic fields.

Future large scale experiments~\cite{proc}, such as the High
Resolution Fly's Eye \cite{Bird4}, the Telescope Array \cite{Teshima},
and most notably the Pierre Auger project \cite{Cronin}, should 
allow to detect clusters of $\ga20$, and possibly more, UHE 
CRs per source, if the clustering suggested by the AGASA
results~\cite{Hayashida1}
is real. The recently proposed satellite observatory
concept for an Orbital Wide-angle Collector (OWL)~\cite{OWL}
might even allow to detect clusters of hundreds of events by
watching the Earth's atmosphere from space. In the present
paper, we wish to examine how
magnetic fields could affect the observations of clusters 
of UHE CRs by future large-scale experiments. We thus assume
that UHE CRs are indeed dominantly protons, and that the magnetic fields
are strong enough to influence their propagation (see below).
We then simulate the 
injection, the propagation, and the detection of UHE CRs
originating from a given source. Finally, we
perform a maximum likelihood analysis on these clusters
of typically $20-50$ particles, and attempt to reconstruct the 
physical parameters describing the source and the magnetic 
fields.
We describe the simulations in Section~2, and discuss the
reconstruction of the different parameters in Section~3; we briefly
summarize our results in Section~4. We use natural units,
$\hbar=c=1$, throughout the paper.

\section{Simulation of clusters of Ultra-High Energy Cosmic
Rays}

Protons of ultra-high energy are subject to the following 
physical processes: energy loss through pair production and
photo-pion production (the latter for $E\ga70\,$EeV) on the cosmic
microwave background, and deflection in the extra-galactic 
magnetic field. In photo-pion production, a proton may be 
converted to a neutron, that either turns back into proton 
through photo-pion production, or decays to a proton on a
distance $\simeq1(E/10^{20}\,{\rm eV})\,$Mpc for a neutron
energy $E$. Pair production is treated as a continuous energy loss 
\cite{CZS}. We treat photo-pion production as a stochastic
energy loss; it is important to do so, as the stochastic nature
of this process imprints significant scatter in arrival time and 
energy for UHE CRs above the GZK cut-off, as discussed in
Ref.\cite{SLO}. We
model the extra-galactic magnetic field as a gaussian random 
field, with zero mean, and a power spectrum given by
$\left\langle B^2(k)\right\rangle\propto k^{n_B}$ for
$k<2\pi/l_c$, and $\left\langle B^2(k)\right\rangle=0$ otherwise.
The cut-off, $l_c$, characterizes the coherence length of the
field. The field is actually calculated on a grid of inter-cell
separation $a$ and is tri-linearly interpolated between the
lattice points such that $l_c\simeq a/\pi$ effectively. The
amplitude of the field is normalized to the r.m.s. strength
$B_{\rm rms}^2\equiv V/(2\pi)^3\int d^3{\bf k}B^2({\bf k})$, and 
our model for the extra-galactic magnetic field is thus 
described by the three parameters $n_B$, $l_c$, and 
$B_{\rm rms}$. Fiducial values for these parameters are
$n_B\simeq0$, $l_c\ga100\,$kpc, and $B_{\rm rms}\la10^{-9}\,$G.
This statistical description of the field allows to treat 
deflection of UHE CRs in the most general case, as discussed in 
Ref.\cite{SLO}. We also note that any relative motion between
observer and source with relative velocity $v$ would introduce
effects only on timescales larger than $l_c/v$ which is much
larger than delay times and experimental lifetimes. It is,
therefore, justified to assume a stationary situation.

The numerical code that we use to follow the propagation of UHE 
protons in an extra-galactic magnetic field is described in 
detail in Ref.\cite{SLO}; here, we summarize its main features.
Protons are injected with a flat energy spectrum, and propagated 
in a given direction in the extra-galactic magnetic field over
a distance $D$, from the source to the detector. Due to the 
stochastic deflection, care has to be taken in how one states 
whether different UHE CRs, that have followed different paths, 
actually reached the same detector, or not \cite{SLO}.
During their propagation, UHE CRs acquire energy-dependent 
deflection $\theta_E$ and time delay $\tau_E$. With a given
sample of nucleons, one can construct different 
histograms, in time, angle, and energy, for different values of 
the differential injection index $\gamma$, and of the
fluence $N_0$. Histograms are then smeared out in energy with  
$\Delta E/E=0.14$, {\it i.e.} a high resolution typical of 
future large-scale UHE CR experiments; histograms are also 
convolved in time with a top-hat of width $T_{\rm S}$, in order 
to simulate emission of particles at the source over a timescale 
$T_{\rm S}$. Once the histogram is obtained for different values 
of the above parameters, clusters of UHE CRs can be obtained by 
picking at random a time window of length
$T_{\rm obs}\simeq5\,$yr, which corresponds to
the lifetime of the experiment, and dialing Poisson statistics over the 
histogram. We do so in order to simulate UHE CR clusters of 
events.

  Conversely, one can use the above code to perform Monte-Carlo 
simulations of UHE CR injection, propagation, and detection, and 
calculating a likelihood of a given histogram for a given 
cluster of events, where the histogram, hence the likelihood, is a
function of the physical parameters described above.
The likelihood is calculated in the standard
way for each observed event cluster, using Poisson statistics,
\begin{equation}
  {\cal L}\left(\tau_{100},T_{\rm S},D,\gamma,
  N_0,l_c,n_B\right)\equiv\left\langle\prod_{j=1,N}
  e^{-\rho_j}\frac{\rho_j^{n(j)}}{n(j){\rm !}}\right\rangle\,,
  \label{likelihood}
\end{equation}
where $\rho_j$ is the predicted number of events in cell $j$, 
and $n(j)$ is the number of observed events
in cell $j$ for the cluster under consideration.
Each cell is defined by a time coordinate and an 
energy. The time-energy histogram is binned to logarithmic
energy bins of size 0.05 in the logarithm to base 10 (as opposed
to 0.1 in Ref.~\cite{SLO} to account for improved energy resolution
of future experiments), and to $0.1\,$yr in linear time bins.
The product in Eq.~(\ref{likelihood}) extends over all energy bins
(from $10^{1.5}\,$EeV to $10^4\,$EeV) and over all time bins
within an observational time window of 
length $T_{\rm obs}$; we took $T_{\rm obs}\simeq5\,$yr as a
projected lifetime of a next generation experiment such as the
Pierre Auger Project. The brackets in Eq.~(\ref{likelihood}) indicate 
that the likelihood has already been averaged with equal weights
over the position of the observational time window on the 
time delay histogram of the UHE CRs, as well as over
different realizations of the extra-galactic magnetic field
between the source and the observer.

The next step is to attempt to reconstruct the parameters in
Eq.~(\ref{likelihood}) from the maximum of the likelihood. 
Future experiments are expected to produce as many as
$\ga100$ particles with $E\ga50\,$EeV, if the AGASA pairs
are real. In the present work, we prefer to remain conservative,
and we simulate clusters of $20-50$ particles with 
$E\ga30\,$EeV.

  In Ref.\cite{LSOS}, we discussed the possible different cases 
of UHE proton images in time, angle and energy, and how, in each 
case, qualitative information could be gained on the magnetic
field and the origin of UHE CRs. Here, we will discuss how each
physical parameter can be reconstructed, and in which case. The 
physical parameters that govern the UHE CRs images are: the time 
delay $\tau_{100}$, normalized at 100EeV,  the coherence length 
$l_c$, the power spectrum index $n_B$, the distance $D$, the 
emission timescale $T_{\rm S}$, the differential injection 
index $\gamma$, and the fluence $N_0$. The time delay is given 
by \cite{WM}:
\begin{equation}
  \tau_E\,\simeq\,
  1.4\,\left(\frac{3+n_B}{2+n_B}\right)
  \left(\frac{D}{30\,{\rm Mpc}}\right)^2
  \left(\frac{E}{100\,{\rm EeV}}\right)^{-2}
  \left(\frac{B_{\rm rms}}{10^{-11}\,{\rm G}}\right)^2
  \left(\frac{l_{\rm c}}{1\,{\rm Mpc}}\right)\;{\rm yr}.
  \label{t_delay}
\end{equation}
Hence, information on $B_{\rm rms}$ is contained in 
$\tau_{100}$. Both the coherence length and the distance play a 
double role. The coherence length not only contributes to the 
time delay, it also influences the scatter around the mean of 
the $\tau_{100}-E$ correlation \cite{WM}: if
$D\theta_E/l_c\ll1$, all UHE CRs have experienced
the same magnetic field 
structure during their propagation, hence the scatter is 
expected to be very small in the absence of pion production;
inversely, if $D\theta_E/l_c\gg1$, 
the scatter is expected to be significant, 
$\Delta\tau_E/\tau_E\sim60$\%, even for negligible energy
loss. The distance also enters the time 
delay, and it also governs the amplitude of pion production, 
hence the high energy part of the spectrum.

  A cluster is seen on the detector as a tri-dimensional image 
in angle, time and energy. As the Monte-Carlo likelihood 
calculation is very time and memory intensive, we only focus on the 
time-energy images in the following. Obviously, information is 
also contained in the angular image itself of the cluster.
For instance, in the limit where $D\theta_E/l_c\ll 1$, one
expects to detect a single image,
albeit shifted by a sytematic offset 
$\theta_E$ from the true location of the source, where
$\theta_E$ is tied to the time delay through 
$\tau_E\simeq D\theta_E^2/4$. Below the GZK cut-off, its
angular size $\Delta\theta/\theta\ll1$.
Note that, provided the cluster
is seen at different energies, and $\theta_E$ is greater than
the angular resolution, the zero-point for $\theta_E$ can be 
reconstructed, as $\theta_E\propto E^{-1}$. In the opposite
limit, $D\theta_E/l_c\gg 1$, one expects the image to be
centered on the source, with an r.m.s. angular size $\theta_E$.
In the intermediate limit, one expects to detect several images.
Moreover, if $\theta_E$ can be measured, it provides an estimate
of the combination $DB_{\rm rms}^2l_c$.

\begin{figure}
\centering\leavevmode
\epsfxsize=6.0in
\epsfbox{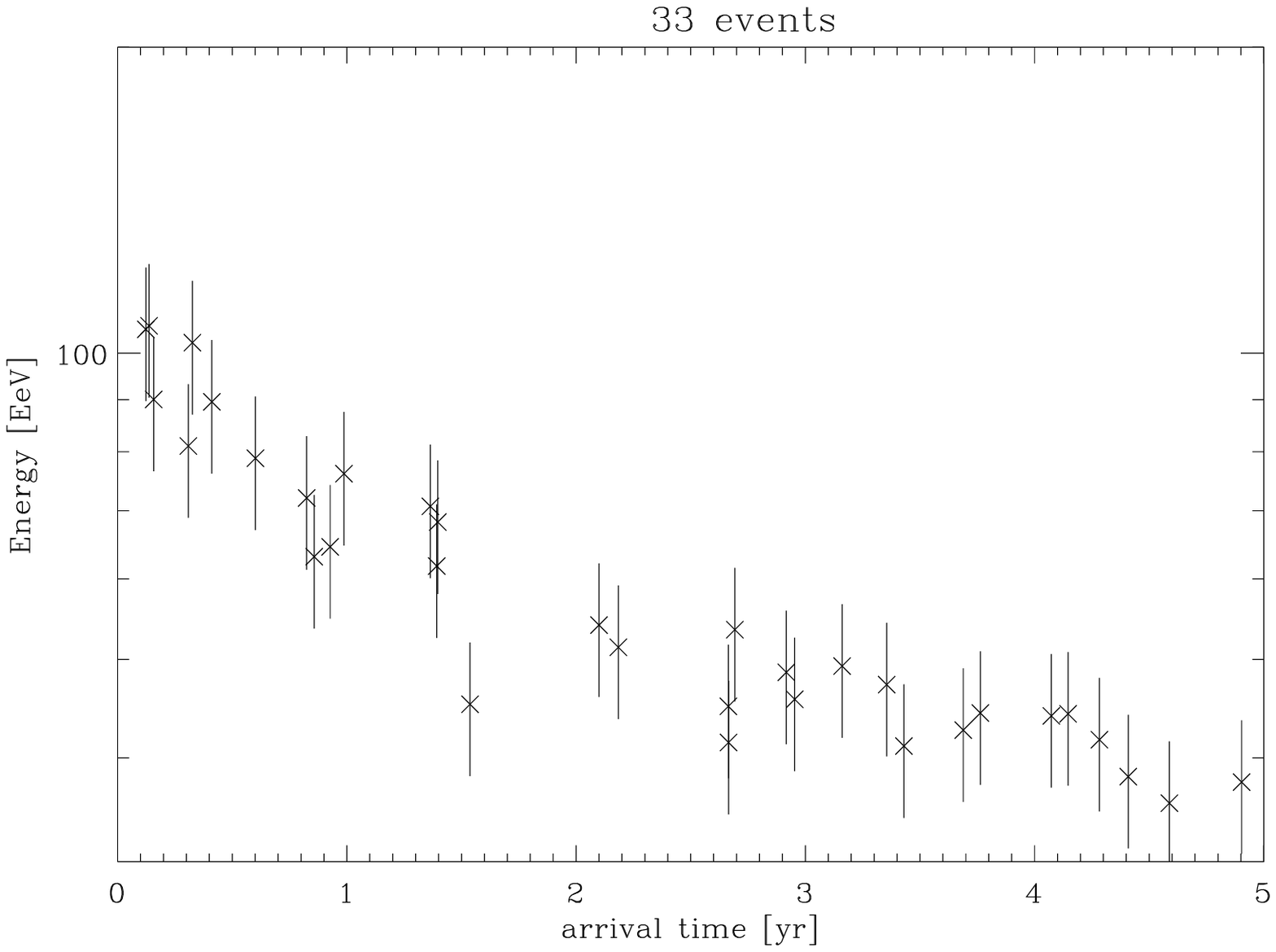}
\caption[...]{A typical cluster of UHE CR above $\simeq30\,$EeV
in the time-energy plane. This cluster was
produced by a discrete source at $D=50\,$Mpc with an emission
time scale $T_{\rm S}\ll1\,$yr ({\it i.e.} a burst), an average
time delay of $\tau_{100}=0.3\,$yr, and with $N_0=40$. For the
extra-galactic magnetic field, a power law index $n_B=0$ and a
coherence length $l_c\simeq1\,$Mpc was assumed. The error bars
correspond to an energy resolution $\Delta E/E=0.14$. The top
label indicates the number of events in the cluster.}
\label{F1a}
\end{figure}

\begin{figure}
\centering\leavevmode
\epsfxsize=6.0in
\epsfbox{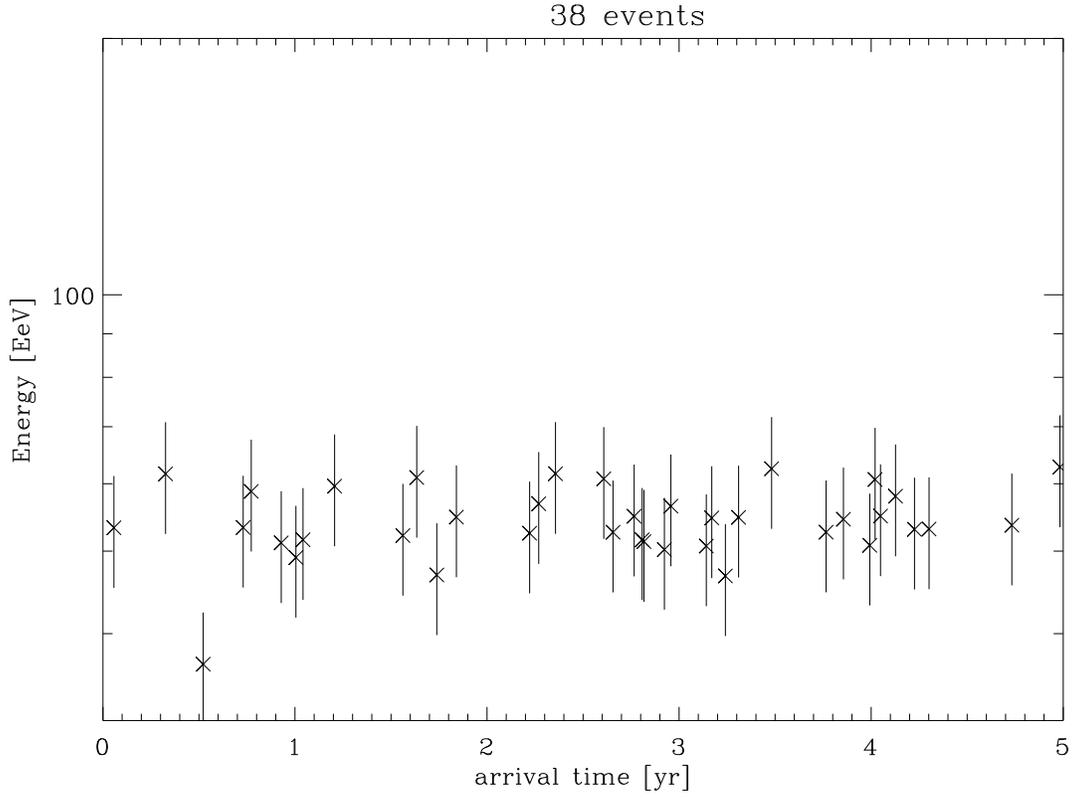}
\caption[...]{Same as for Fig.~\ref{F1a}, except for the
parameters $\tau_{100}=50\,$yr, and $N_0=4\times10^3$. This serves as
an example for a
burst with a long time delay, but still in the limit of small
deflection, $D\theta_E/l_c\ll1$, leading to a small detected
energy dispersion.}
\label{F1b}
\end{figure}

\begin{figure}
\centering\leavevmode
\epsfxsize=6.0in
\epsfbox{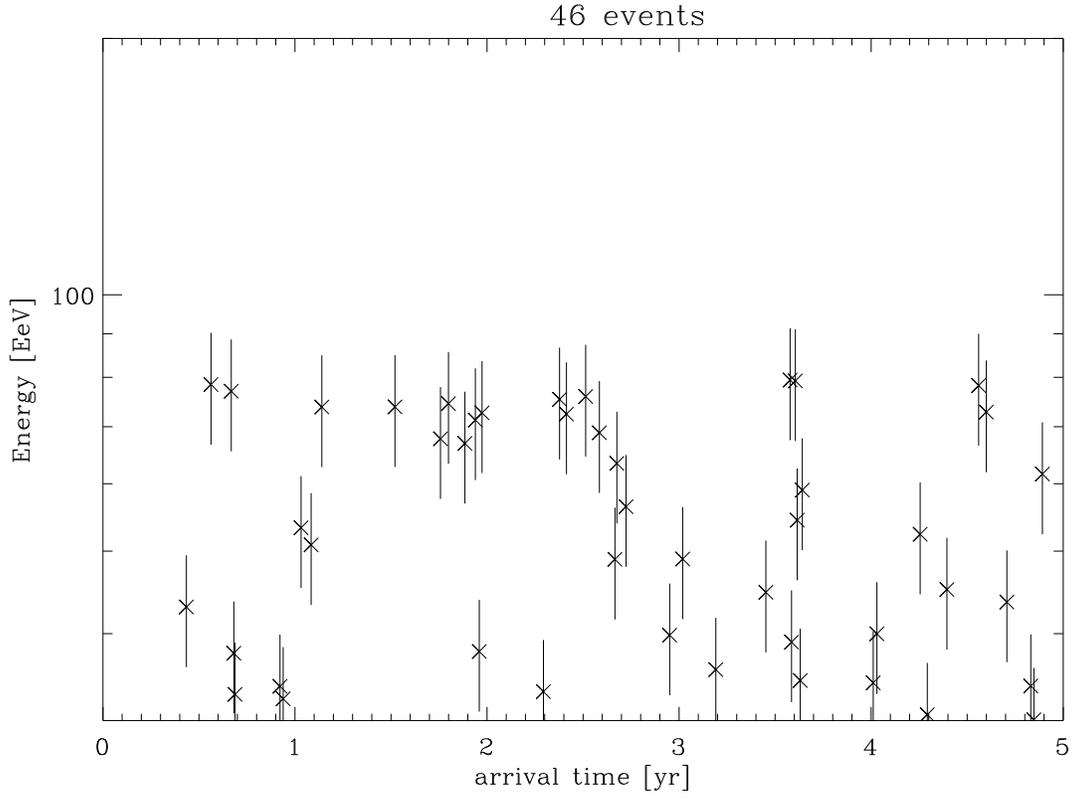}
\caption[...]{Same as Fig.~\ref{F1a}, except for the parameters
$\tau_{100}=250\,$yr, $N_0=2\times10^4$, and
$l_c\simeq0.25\,$Mpc. This serves as an
example for a burst with a long time delay for intermediate
deflection, $D\theta_E/l_c\sim1$. The distinct sub-bands are due
to multiple source images.}
\label{F1c}
\end{figure}

\begin{figure}
\centering\leavevmode
\epsfxsize=6.0in
\epsfbox{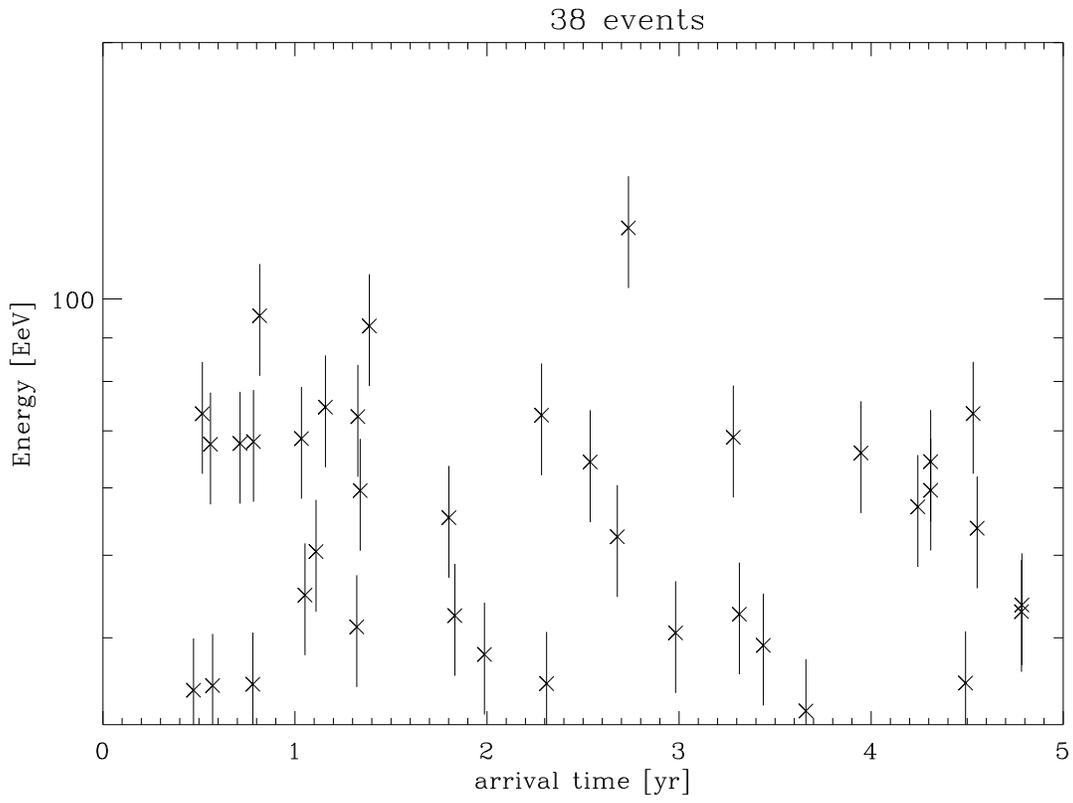}
\caption[...]{Same as Fig.~\ref{F1a}, except for the parameters
$\tau_{100}=0.1\,$yr, $T_{\rm S}=500\,$yr, and $N_0=4\times10^3$. This
serves as an example of a source that is continuously emitting
at all relevant energies.}
\label{F1d}
\end{figure}

\begin{figure}
\centering\leavevmode
\epsfxsize=6.0in
\epsfbox{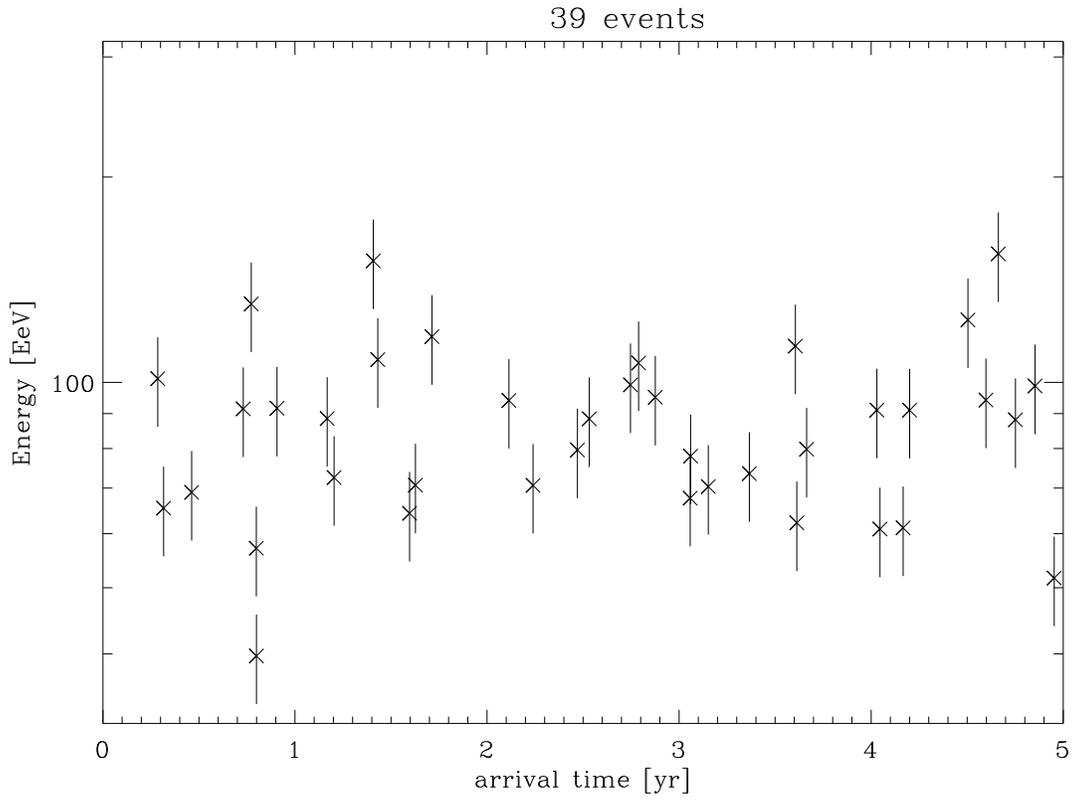}
\caption[...]{Same as Fig.~\ref{F1a}, except for the parameters
$\tau_{100}=50\,$yr, $T_{\rm S}=200\,$yr, and
$N_0=6\times10^3$. A lower cut-off in
energy occurs at $E_{\rm C}\simeq50\,$EeV where $\tau_{E_{\rm
C}}\simeq T_{\rm S}$.}
\label{F1e}
\end{figure}

  Main features of the time-energy images of clusters of UHE 
protons are described in detail in Ref.\cite{LSOS}. We summarize 
these results briefly, as they are important to the following. 
If both $T_{\rm S}<\tau_{100}$, and $\tau_{100}$ is small
compared to
$T_{\rm obs}$, arrival time and energy are correlated according
to $\tau_E\propto E^{-2}$; see Fig.~\ref{F1a}. A source, such that 
$\tau_{100}\gg T_{\rm S}$ and $\tau_{100}\gg T_{\rm obs}$, can 
be seen only in a limited range of energies, at a given time, as
discussed in Ref.\cite{WM}, as shown in Fig.~\ref{F1b},\ref{F1c}.
Below the GZK cut-off, the width of
this stripe, in the time-energy plane and within the
observational window of length $T_{\rm obs}$, is then tied to
the ratio $D\theta_E/l_c$, as discussed above.
At the other extreme, a source emitting continuously at all
energies of interest here, {\it i.e.} with $T_{\rm S}\gg\tau_{30}$ and
$T_{\rm S}\gg T_{\rm obs}$, yields a time-energy image
in which the distribution of arrival time
{\it vs.} energy is uniform, {\it i.e.} events of any energy can
be recorded at any time, as shown in Fig.~\ref{F1d}.
Finally, for a source, such that $\tau_{100}< T_{\rm S}$ and
$\tau_{30}> T_{\rm S}$, together with $T_{\rm S}\gg T_{\rm obs}$,
there exists an energy $E_{\rm C}$, such 
that $\tau_{E_{\rm C}}=T_{\rm S}$. In this case, protons with an 
energy lower than $E_{\rm C}$ are not detected, as they could 
not have reached us within $T_{\rm obs}$, even if they were 
among the first emitted. However, protons with an energy higher
than $E_{\rm C}$ are detected as for a continuously emitting
source, {\it i.e.} with a uniform distribution of arrival times
{\it vs.} energy, see Fig.~\ref{F1e}.

Typical simulated clusters corresponding 
to these five main situations are shown in 
Figs.~\ref{F1a}-\ref{F1e}. The fluence $N_0$ was normalized in each 
case so that $\simeq40$ events are expected within 5 years.

\section{Maximum likelihood reconstruction}

In this section, we discuss, in turn, how each parameter can be 
obtained from a likelihood study of UHE CR clusters. Certain
marginalizations of Eq.~(\ref{likelihood}) are used whenever the
focus is only on one or a part of the parameters. The other
parameters are then averaged or integrated over, applying weight
functions ({\it i.e.}, Bayesian priors) that represent the prior
knowledge on their values. As we have currently no
information on the fluence, the emission timescale $T_{\rm S}$ and
the time delay $\tau_E$, the prior chosen would be uniform in the
logarithm of these parameters. However, we note that the time delay
$\tau_E$ is bounded from above by the Faraday rotation data bound on 
$B_{\rm rms}l_c^{1/2}$ as combined with Eq.~(\ref{t_delay}). Moreover,
information contained in the angular image should also be included in
the prior on $\tau_E$, as $\tau_E\simeq D\theta_E^2/4$. The 
marginalization over the injection spectral index $\gamma$ is achieved
through averaging with equal weights.

Although we focus on only one source, future large-scale
experiments are expected to detect a large number of individual
sources. Obviously, this would considerably increase the sensitivity
to the physical parameters.

\subsection{Time delay $\tau_E$}

  Here we assume that the source is a burst, {\it i.e.} 
$T_{\rm S}\ll 1\,$yr; we will discuss
the case where $T_{\rm S}\gg1\,$yr in the section concerning
$T_{\rm S}$.

  If the time delay is small compared to the length of the 
observational window, the time-energy correlation is scanned 
through, and, as Fig.~\ref{F1a} reveals, as simple fit of 
$\tau_E\propto E^{-2}$ would allow to determine the zero-point 
of emission, hence the time delay. This constitutes a
measurement of the combination
$DB_{\rm rms}l_c^{1/2}$. Our likelihood simulations confirm 
that, for the cluster shown in Fig.~\ref{F1a} for instance, 
$\tau_E$ is obtained within a factor 2. The source 
is found to be a burst with a high level of confidence.

\begin{figure}
\centering\leavevmode
\epsfxsize=6.0in
\epsfbox{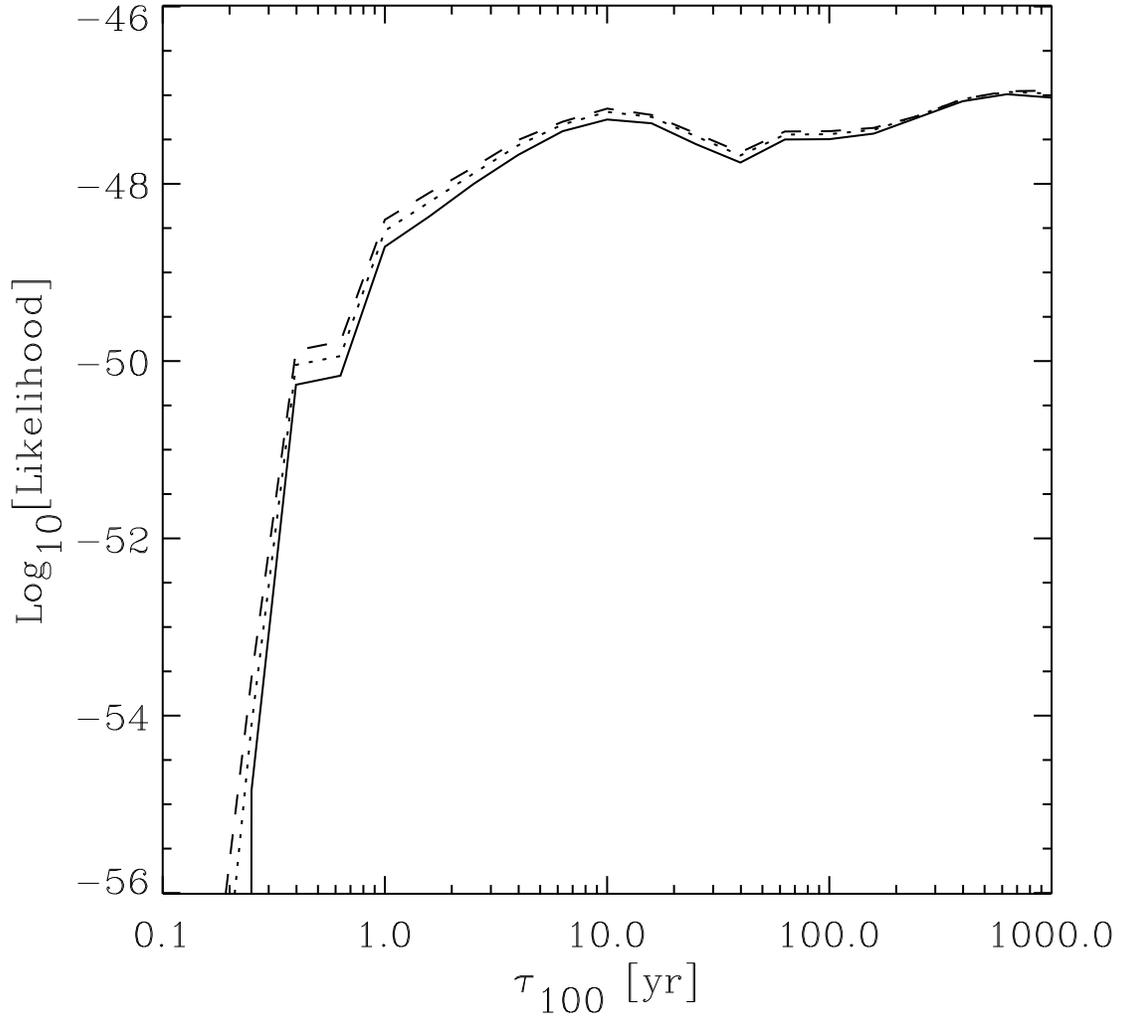}
\caption[...]{The likelihood for the cluster shown in
Fig.~\ref{F1b}, marginalized over $T_{\rm S}$ and $N_0$,
plotted versus the time delay $\tau_{100}$, for $D=50\,$Mpc and
$l_c\simeq1\,$Mpc (the true values). The solid line is
for $\gamma=1.5$, the dotted for $\gamma=2.0$ (the true value),
and the dashed line for $\gamma=2.5$. As a result,
$\tau_{100}>5\,$yr to about 90\% confidence level. In this case
there is clearly no sensitivity to $\gamma$, as expected.}
\label{F2}
\end{figure}

  When the time delay gets significantly larger than
$T_{\rm obs}$, its actual value cannot be reconstructed 
from the maximum of the likelihood. This case
corresponds to the clusters shown in Figs.~\ref{F1b} and~\ref{F1c}.
Indeed, the likelihood is degenerate in the parameters $N_0$ and 
$\tau_{100}$, as it depends mainly on the rate of detection
$N_0/\Delta\tau_{100}$, where $\Delta\tau_{100}$ is the scatter
in time around the mean of the $\tau_E-E$ correlation. 
As long as $N_0$ is unknown, only a lower
limit to $\tau_{100}$, typically $\tau_{100}\ga T_{\rm obs}$, can
be placed. The likelihood, as calculated for the cluster shown in 
Fig.~\ref{F1b}, and marginalized over all parameters except 
$\tau_{100}$ and $\gamma$, is shown in Fig.~\ref{F2} in order to
illustrate this point. The distance and the coherence length
cannot be readily obtained in this case, as we will discuss further
below. Although only a lower 
limit could be placed on the time delay, we note that, 
when combined with the Faraday rotation bound, this would still allow
to bracket the strength of the extra-galactic magnetic field, within 
less than a few orders of magnitude. 

At this point, the information 
contained in the angular image of the source becomes important. If the 
angular image is not resolved, this translates into an upper limit on 
$\tau_E/D$, which may supersede the Faraday rotation bound, see 
Eq.~(\ref{t_delay}), and Eq.~(\ref{ang}) below. At the other extreme,
for a sufficiently large time delay, $\theta_E$ should in principle
be measurable, as
\begin{equation}
  \theta_E\simeq0.02^{\circ}\left(\frac{D}{10\,{\rm Mpc}}\right)^{-1/2}
  \left(\frac{\tau_E}{1\,{\rm yr}}\right)^{1/2}\,.\label{ang}
\end{equation}

Obviously, resolving the angular image would change the prior for $D$ 
and $\tau_{100}$; it would sharpen the maximum likelihood 
reconstruction, notably with respect to the various scenarios
discussed in Section~2. We have not included this angular effect  
in a systematic way; a quantitative treatment of the angular images
will be the subject of a future study. We note that the angular 
resolutions of future UHE CR experiments are fractions of a degree, 
hence the information contained in the angular image becomes 
significant for $\tau_{100}\ga 10\,$yr.

\subsection{Distance $D$}

  As mentioned above, the distance enters the likelihood mainly 
through the amplitude of pion production. As long as the 
high energy tail of the spectrum, {\it i.e.} $E\ga50\,$EeV,
can be observed, the distance is thus obtained with a reasonably good
accuracy from the likelihood, as marginalized over $T_{\rm S}$,
$\tau_{100}$, $N_0$, and $\gamma$. In particular, the likelihood is
sensitive to the distance if the source has a large emission
timescale, $T_{\rm S}\gg\tau_{100}$. The standard error
is then roughly a
factor $\simeq 2$. For example, the cluster of Fig.~\ref{F1d}
shows a factor $\simeq5$ difference in the marginalized
likelihood for $50\,$Mpc (the true value) and $30\,$Mpc; such a 
factor is a typical value. The difference between $50\,$Mpc and
$5\,$Mpc is typically a factor $\simeq15$.

If $T_{\rm S}\ll\tau_{100}\la
T_{\rm obs}$, and the range of energies seen by 
the detector is above the GZK cut-off, the distance can still be 
evaluated, albeit with a somewhat larger error. Typical
differences in the likelihood between 50 and $30\,$Mpc and 50
and $5\,$Mpc are factors $\simeq2$ and $\simeq6$,
respectively.

In the intermediate case where $\tau_{100}$ and
$T_{\rm S}$ are comparable, so that $\tau_{E_{\rm C}}=T_{\rm S}$
for an $E_{\rm C}$ in the observable energy range, the sensitivity to
$D$ is the better the lower $E_{\rm C}$, albeit not very strong. The
difference in the likelihood between $50\,$Mpc and $30\,$Mpc is
typically a factor 3 or less ({\it e.g.}, for the cluster shown in
Fig.~\ref{F1e}). It quickly rises, however, to a factor $\simeq20$
for clusters of the order of 100 events.
We note that, due to the comparatively limited energy range seen
in this case, there is a partial degeneracy
between $D$ and the injection spectrum parametrized by
$\gamma$. For example, the marginalized likelihood does not
change significantly if $D$ is decreased and $\gamma$ is increased
({\it i.e.} a softer injection spectrum is assumed) at the same time.

Other cases, {\it e.g.}, as shown in Figs.~\ref{F1b}
and~\ref{F1c}, do not allow to reconstruct $D$.

\subsection{Emission timescale $T_{\rm S}$}

  If the emission timescale is larger than the width of the 
observational window, the likelihood becomes
degenerate in the ratio $N_0/T_{\rm S}$, and only a lower limit 
to $T_{\rm S}$ can be obtained, typically
$T_{\rm S}\ga T_{\rm obs}$.
However, if the time delay at some intermediate energy, between 
say $30\,$EeV and $100\,$EeV, is sufficiently large, and comparable
to the emission timescale, then both the time delay, and the
emission timescale, can be measured as long as a lower energy
cut-off is visible above which the emission appears
continuous. This case corresponds to 
the cluster shown in Fig.~\ref{F1e}. Indeed, if the time delay 
is sufficiently large, then $\theta_E$ can be observationally 
measured according to Eq.~(\ref{ang}).
As discussed above, the likelihood has some sensitivity to the
distance as long as events are observed over a reasonable range of
energies. This sensitivity also depends strongly on the statistics of 
the UHE CR cluster. Since
$\tau_E\simeq D\theta_E^2/4$, $\tau_E$ is obtained, and, as 
discussed in Ref.\cite{LSOS}, the emission timescale corresponds 
to the time delay at the cut-off energy $E_{\rm C}$, below which
no UHE CR are recorded within $T_{\rm obs}$, as follows from the
definition of this lower cut-off energy,
$\tau_{E_{\rm C}}=T_{\rm S}$.

In reality, however, the image observed in such a situation will
appear as a burst with a large time delay most of the time:
for $E<E_{\rm C}$, the image is similar to that of a burst with 
a large time delay, as $\tau_E> T_{\rm S}\gg T_{\rm obs}$, {\it 
i.e.} only a limited range in energies is detected. Because
$\tau_{30}>\tau_{E_{\rm C}}$, most sources are seen at
$E<E_{\rm C}$ rather than at $E>E_{\rm C}$, where the image is
similar to that of a continuous source. Notably, the likelihood 
for a bursting source with $T_{\rm S}\simeq0$ does not exclude
the above intermediate situation, for a cut-off $E_{\rm C}$ above
the observed stripe in the time-energy image of a bursting source.
In the case of Fig.~\ref{F3}, corresponding to the cluster shown in
Fig.~\ref{F1c}, the  stripe is observed between $\simeq30\,$EeV and 
$\simeq80\,$EeV, and the likelihood does not exclude the above
intermediate case with $E_{\rm C}\ga80\,$EeV.
Needless to say, the  best reconstruction of $\tau_{100}$ and
$T_{\rm S}$ takes place when the source is observed above
$E_{\rm C}$, see Fig.~\ref{F4}.

  Finally, note that if  $T_{\rm S}\gg\tau_{30}$, the 
continuous source is hardly mistaken for a burst with a large 
time delay, which would be the closest approximation to the 
time-energy image of a continuous source. This can be seen in
Fig.~\ref{F5a}, which represents contours of the likelihood in
the $\tau_{100}-T_{\rm S}$ plane. If the likelihood is further 
marginalized with respect to $T_{\rm S}$ or $\tau_{100}$ (see
Fig.~\ref{F5b}), a burst with a large time delay is ruled out to
about 95\% confidence level.
Qualitatively speaking, the difference is that for a
burst with a large time delay, the maximum fluence occurs at some 
intermediate energy, and the fluence decreases with decreasing 
energy below. For a continuous source, in contrast, the fluence 
increases with decreasing energy, according to the injection 
negative power law spectrum.

\subsection{Injection spectrum index $\gamma$}

  The injection spectrum index $\gamma$ can be measured provided 
UHE CRs are recorded over a bandpass in energy that is 
sufficiently broad. More precisely, in the case of a continuous 
source, {\it i.e.} $T_{\rm S}\gg\tau_{30}$, $\gamma$ can be 
measured with an absolute accuracy of $\simeq0.3$. This is based
on the likelihood as marginalized over $T_{\rm S}$, $\tau_{100}$,
and $N_0$, albeit for a known distance $D$. For example,
for a continuously emitting source at $D=50\,$Mpc with
$\gamma=2.0$ (see, {\it e.g.}, Fig.~\ref{F1d}), we
obtained a difference in the likelihood for $\gamma=1.5$ and 2.5
of a factor of about 30 and 2, respectively, on average. An
example for this situation is given in Fig.~\ref{F5b}.
In the case of a continuous source with a time delay comparable
to the emission timescale, {\it i.e.} such as shown in Fig.~\ref{F1e},
the respective factors are about 2 and 1, and
therefore hardly significant. For a burst with a small time
delay such as in Fig.~\ref{F1a} these factors are about 10 and
1. A burst with $\tau_{100}\gg T_{\rm obs}$
in which case the signal would be spread over a large 
range in energy, is even less sensitive to $\gamma$. In general,
therefore, it is comparably easy to rule out a hard injection
spectrum if the actual $\gamma\ga2.0$, but it is much harder
to distinguish between $\gamma=2.0$ and 2.5.

Our analysis of the sensitivity to $\gamma$ was restricted to a
fixed distance $D$, mainly because of CPU time limitations of
the present serial version of our code. We expect that in the
absence of information on $D$, an additional marginalization
over $D$ would decrease the sensitivity to $\gamma$. In particular,
we already mentioned at the end of Section~3.2, a degeneracy of
the likelihood between $\gamma$ 
and $D$ for the intermediate case, where $T_{\rm S}$ and
$\tau_{E_{\rm C}}$ are comparable.

\subsection{Fluence $N_0$}

Because of the degeneracy of the likelihood in $N_0/T_{\rm S}$
and/or $N_0/\Delta\tau_{100}$ for large timescales, it is in general
not possible to reconstruct $N_0$. A possible exception is the
case where all the particles are detected, {\it i.e.} 
$\tau_{100}\la T_{\rm obs}$, and $T_{\rm S}\la T_{\rm
obs}$. 

We take advantage of this section to detail slightly the 
marginalization procedure over $N_0$.
In most cases we marginalized over the fluence analytically,
noting that the dependence of the likelihood on $N_0$ can be
written as
\begin{equation}
  \ln{\cal L}=a\exp(x)+bx+c\,,\label{margN0}
\end{equation}
with $x\equiv\ln N_0$ and where $a$, $b$, and $c$ depend on all
other parameters except $N_0$. This just follows from the fact that
$\rho_j$ in Eq.~(\ref{likelihood}) is proportional to $N_0$. By
using the approximation
\begin{equation}
  \ln{\cal L}(x)\simeq\ln{\cal L}_{\rm max}-
  \frac{b}{2}(x-x_{\rm max})^2\label{N)approx}
\end{equation}
in terms of value ${\cal L}_{\rm max}$ and location $x_{\rm
max}$ of the $N_0-$maximized likelihood, marginalizing over
$N_0$ with a uniform prior for $\ln N_0$ then amounts to
computing
\begin{equation}
  \int_{-\infty}^{+\infty}dx{\cal L}(x)=
  {\cal L}_{\rm max}\left(\frac{2\pi}{b}\right)^{1/2}
  \,.\label{N0int}
\end{equation}

\subsection{Coherence length $l_c$}
  
\begin{figure}
\centering\leavevmode
\epsfxsize=6.0in
\epsfbox{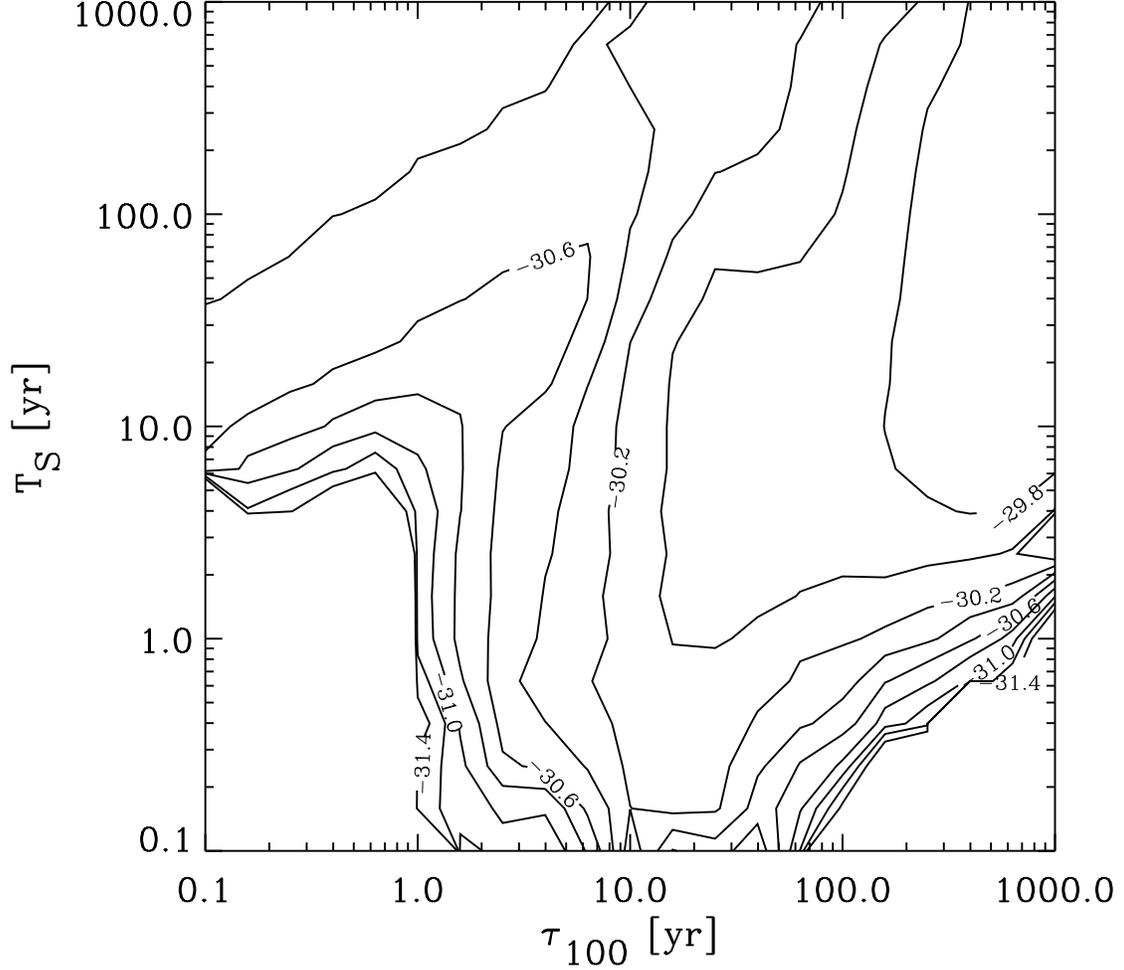}
\caption[...]{The likelihood for the cluster shown in
Fig.~\ref{F1c}, marginalized over $N_0$ and $\gamma$, plotted
in the $\tau_{100}-T_{\rm S}$ plane, for $D=50\,$Mpc and
$l_c\simeq0.25\,$Mpc (the true values). The maximum of the
likelihood occurs for $\tau_{100}\simeq$ a few hundred years,
$T_{\rm S}\simeq$ a few years which
is a good reconstruction of the true values. The contours shown
go from the maximum down to values of about 0.01 of the maximum
in steps of a factor $10^{0.2}=1.58$.
Note that values in the range $\tau_E=T_{\rm S}$ with
$E\ga80\,$EeV and $T_{\rm
S}\ga10\,$yr are not significantly excluded, as expected
(see text). The fall-off at $\tau_{100}\ga50\,$yr and $T_{\rm
S}\la3\,$yr is a numerical artifact due to the limited number
of propagated particles ($4\times10^4$ per parameter set) which
causes too patchy histograms in arrival time.}
\label{F3}
\end{figure}

\begin{figure}
\centering\leavevmode
\epsfxsize=6.0in
\epsfbox{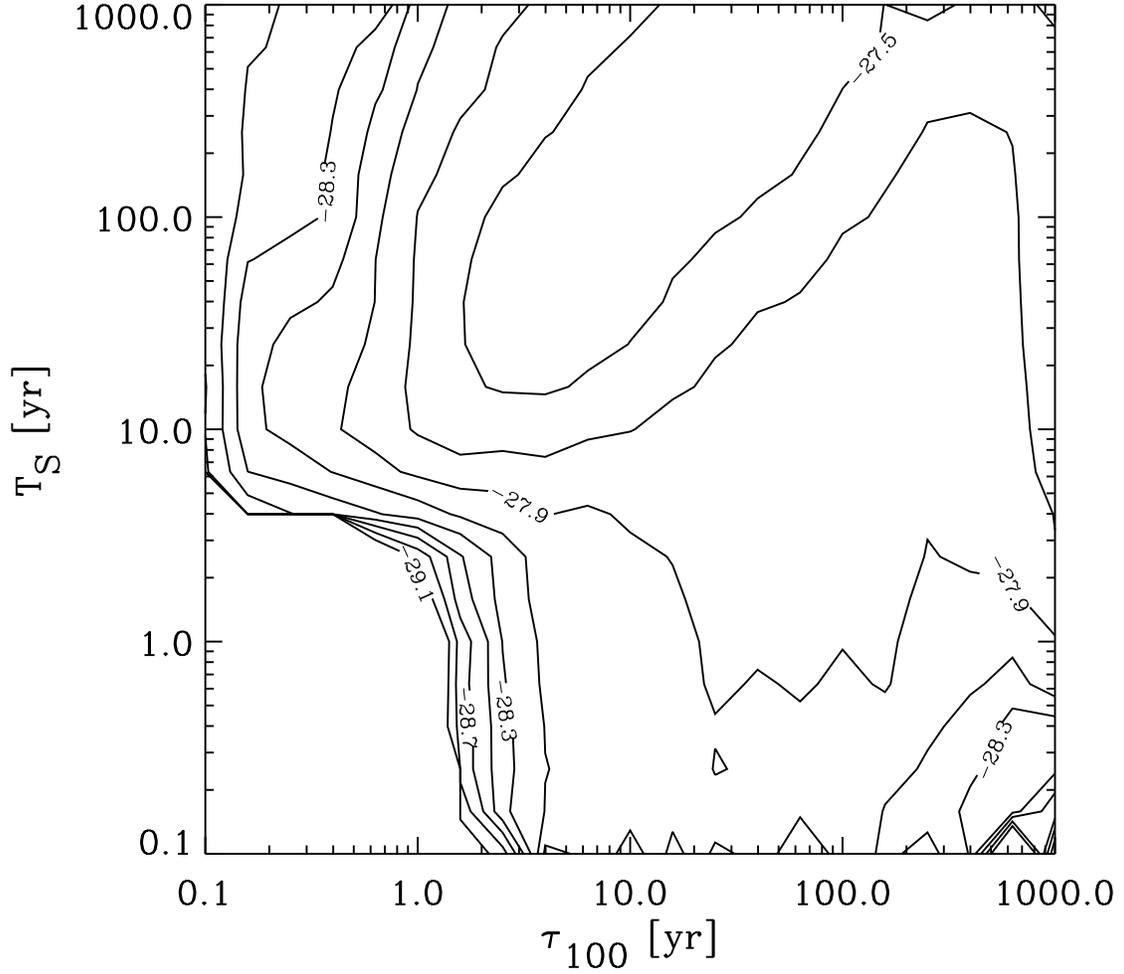}
\caption[...]{Same as Fig.~\ref{F3}, but for the cluster shown
in Fig.~\ref{F1e}, and assuming $l_c\simeq1\,$Mpc (the true
value). The maximum of the likelihood is near
$\tau_{100}=4\,$yr, $T_{\rm S}=100\,$yr, but any values along
the ridge defined by
$T_{\rm S}\simeq\tau_{50}$ are roughly equally likely, as
expected from the fact that $E_{\rm C}\simeq50\,$EeV (see
Fig.~\ref{F1e} and text).}
\label{F4}
\end{figure}

\begin{figure}
\centering\leavevmode
\epsfxsize=6.0in
\epsfbox{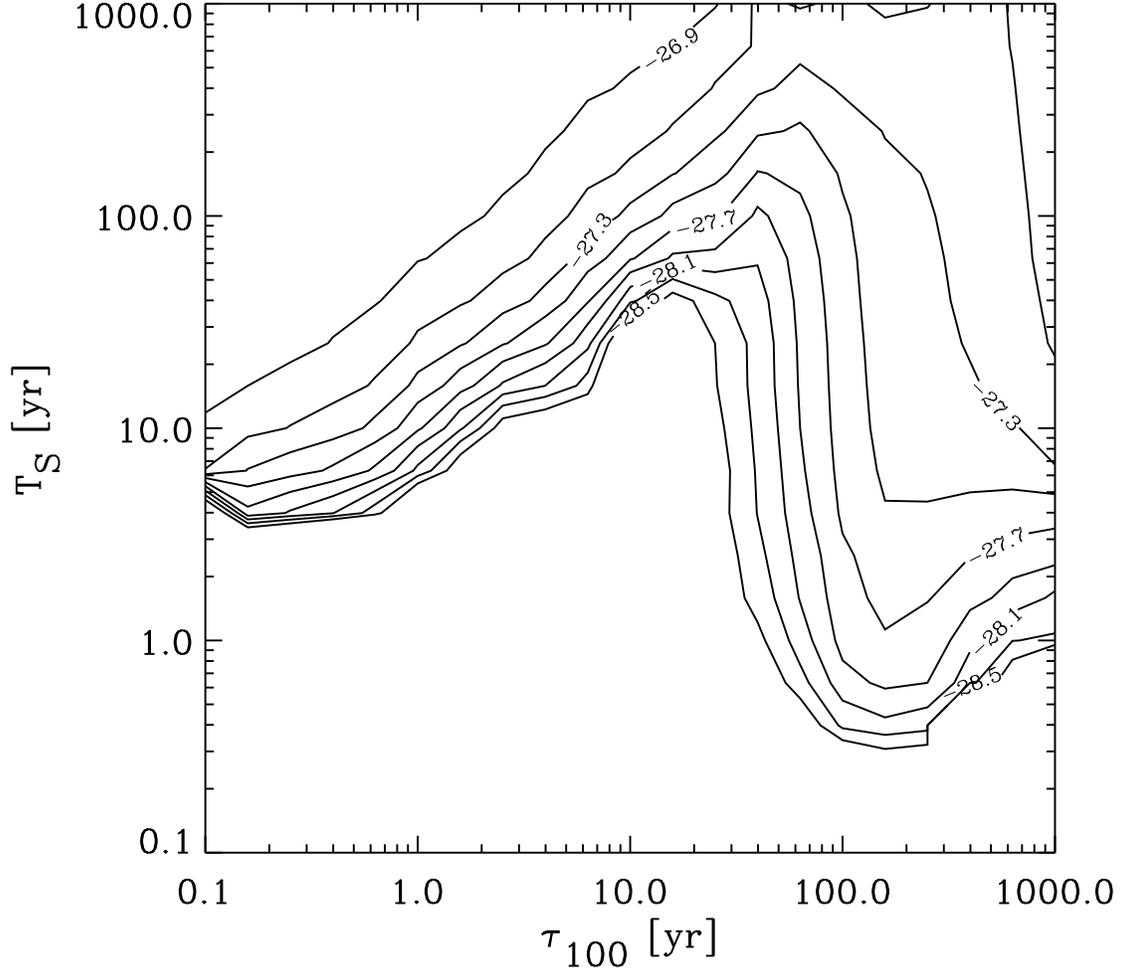}
\caption[...]{Same as Fig.~\ref{F3}, but for the cluster shown
in Fig.~\ref{F1d}, and assuming $l_c\simeq1\,$Mpc (the true
value). The maximum of the likelihood is near
$\tau_{100}=0.1\,$yr, $T_{\rm S}=10^3\,$yr which
is a good reconstruction of the true values. Parameters in the
range $T_{\rm S}\la10\,$yr and $\tau_{100}\ga10\,$yr,
that would be typical for a burst with a large time delay, are
excluded at about 95\% confidence level (see text).}
\label{F5a}
\end{figure}

\begin{figure}
\centering\leavevmode
\epsfxsize=6.0in
\epsfbox{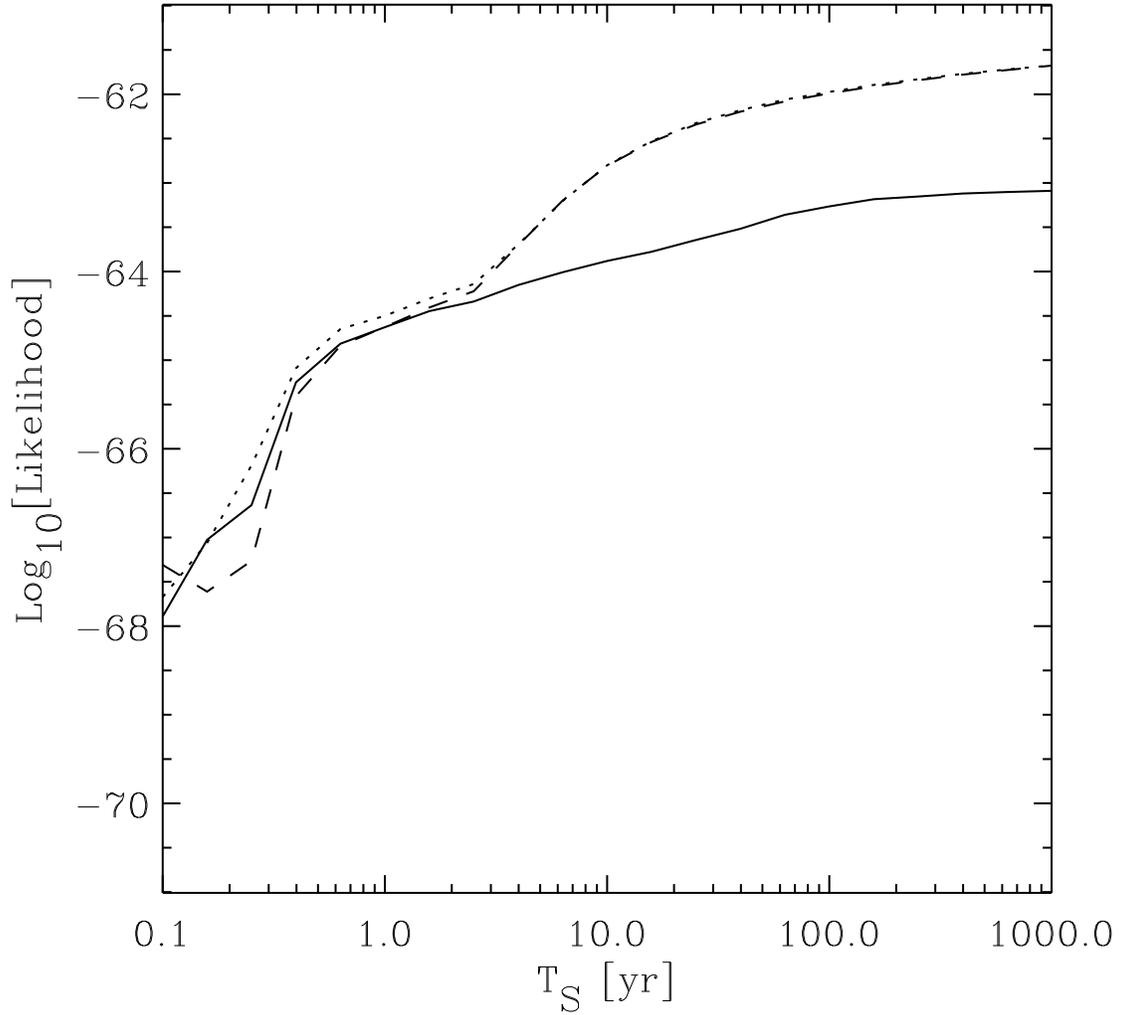}
\caption[...]{The likelihood for the cluster shown in
Fig.~\ref{F1d}, marginalized over $\tau_{100}$ and $N_0$,
plotted versus the emission timescale $T_{\rm S}$, for
$D=50\,$Mpc and $l_c\simeq1\,$Mpc (the true values). The average over
the curves shown for $\gamma=1.5$ (solid line), $\gamma=2.0$
(dotted line; the true value),
and for $\gamma=2.5$ (dashed line), therefore, corresponds
to a marginalization of the likelihood shown in Fig.~\ref{F5a}
over $\tau_{100}$. This demonstrates even more clearly that
$T_{\rm S}\la10\,$yr is ruled out at about 95\% confidence
level.}
\label{F5b}
\end{figure}

\begin{figure}
\centering\leavevmode
\epsfxsize=6.0in
\epsfbox{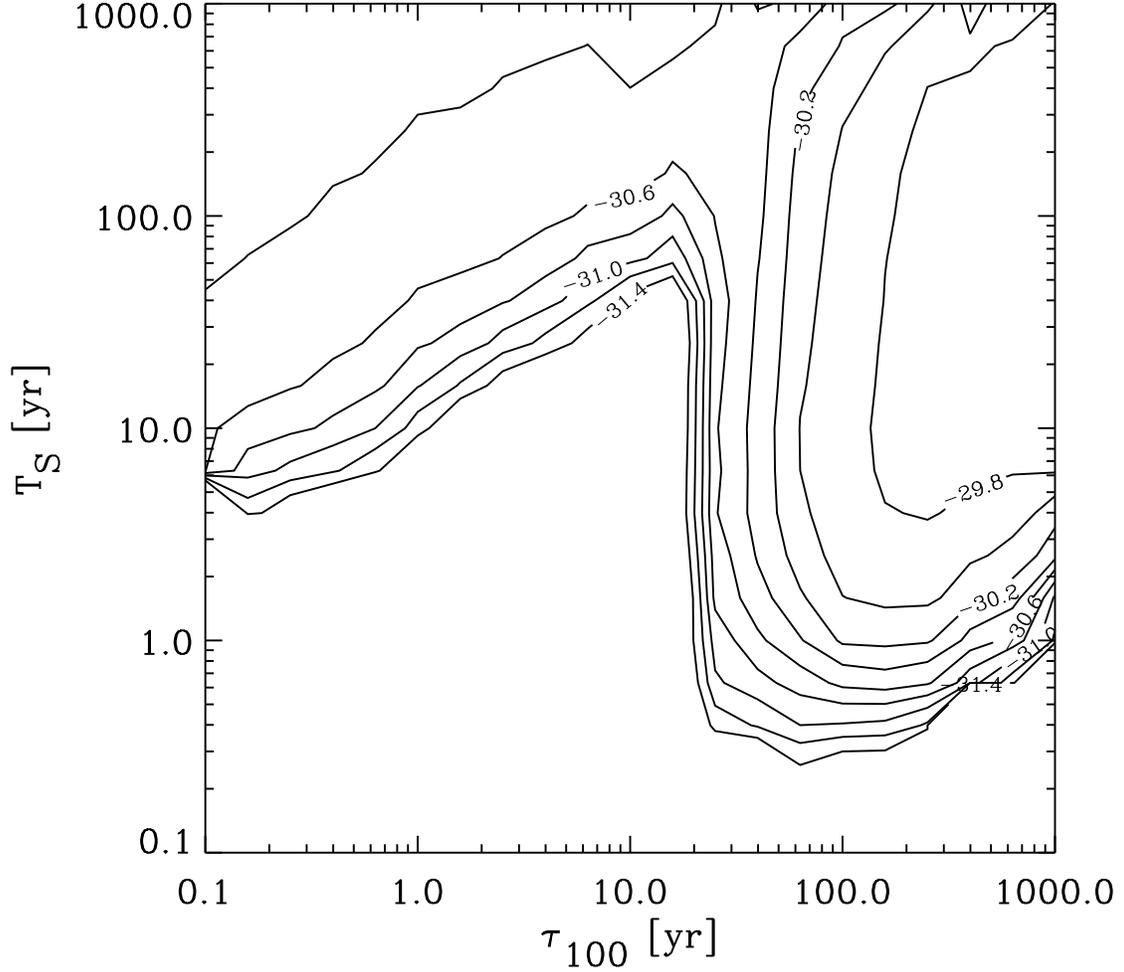}
\caption[...]{Same as Fig.~\ref{F3}, but assuming a magnetic
field coherence length $l_c\simeq1\,$Mpc. The maximum of the
likelihood is near $\tau_{100}=400\,$yr, $T_{\rm S}=20\,$yr
which again is a reasonable reconstruction of the true values.
Note that if $\tau_{100}$ were known to be smaller than
$\simeq50\,$yr, a coherence length as large as $l_c=7\,$Mpc
could be ruled out, but $l_c\simeq0.25\,$Mpc would be allowed (see
Fig.~\ref{F3} and discussion in the text).}
\label{F6}
\end{figure}

  Our simulations confirm the suggestion of Ref.\cite{WM}, that the 
main effect of $l_c$ on the angle-time-energy image comes through the 
relative size of the scatter around the $\theta_E-\tau_E-E$ 
correlations. For a time-energy image, if the source is continuous,
with $T_{\rm S}\gg\tau_{30}$, the correlation between $\tau_E$ and
$E$ is drowned in the uniform emission of particles within the
timescale $T_{\rm S}$, and $l_c$ plays no role. The coherence 
length can therefore be estimated only when
$\tau_{100}\gg T_{\rm S}$, and $\tau_{100}\gg T_{\rm obs}$. In 
this case, the signal corresponds to that shown in
Figs.~\ref{F1b} and~\ref{F1c}, and the width of the signal is
related to the value of $D\theta_E/l_c$. The
likelihood marginalized over $N_0$ and $\gamma$ for the cluster
Fig.~\ref{F1c}, assuming two different coherence lengths,
$l_c\simeq0.25\,$Mpc and $l_c\simeq1\,$Mpc, is shown in Figs.~\ref{F3}
and~\ref{F6}. The qualitative behavior of these contour plots is the 
following. The actual value of $l_c$ used in simulating the cluster is 
$l_c\simeq0.25\,$Mpc, so that the likelihood shown in Fig.~\ref{F3} uses the 
correct value of $l_c$. The width of a stripe in the time-energy 
plance, is tied to the ratio $D\theta_E/l_c$, or, equivalently, to
$(D\tau_E)^{1/2}/l_c$. The likelihood shown in Fig.~\ref{F6}, 
assuming $l_c\simeq1\,$Mpc, is thus similar to that corresponding to 
$l_c\simeq0.25\,$Mpc, although it requires a comparatively larger
({\it i.e.} $\simeq(1/0.25)^2$ times larger) time delay to reproduce
the large scatter in the stripe. 

This demonstrates that for a broad observed energy
dispersion, a large coherence length can be ruled out at least
when some information on the distance $D$ and the deflection angle
$\theta_E$, and thus on $\tau_E$, is available. In contrast,
ruling out a small coherence length for a small observed energy
dispersion is much harder, due to the nature of Poisson statistics.

  As mentioned previously, provided 
$\tau_E\ga10^4\,$yr, $\theta_E$ can be directly measured. Note 
that the upper limit on the magnetic field strength, obtained 
through Faraday rotation measurements, implies
$\tau_E\la 2\times10^5\,{\rm yr}(D/10\,{\rm Mpc})^2(E/10\,{\rm
EeV})^{-2}$. Whenever $\theta_E$ is measurable, the degeneracy
of the likelihood,
with respect to the width of the signal in energy, thus 
concerns only the ratio
$D/l_c$. As we argued above, the likelihood itself is not 
sensitive to the distance if the mean energy of the signal lies 
below the GZK cut-off. If the mean energy lies above the GZK 
cut-off, then not only $\theta_E$ should not be measurable,
but the width itself of the signal now arises from two very 
different effects: one due to the different trajectories 
followed by different UHE CRs through the magnetic field, another 
due to the pion production stochastic broadening of the 
signal; these effects cannot be easily disentangled. Hence, the only 
way this degeneracy between $D$ and $l_c$ could be broken is 
through the observation of an astrophysical counterpart to the 
source, or the source host, and the direct measurement of its 
distance.

Finally, the likelihood was found extremely insensitive to the index 
$n_B$ of the power spectrum of magnetic inhomogeneities.

\section{Conclusions}

  We have presented a Monte-Carlo likelihood analysis of the potential 
of future large-scale UHE CR experiments to reconstruct the 
physical parameters of the source and of intervening magnetic 
fields, when the strength of the latter is sufficient ($B_{\rm 
rms}\ga10^{-12}\,$G) to affect the propagation of UHE CRs.

  We discussed five generic situations of the time-energy images of 
UHE CRs, which we classify according to the values of the time delay 
$\tau_E$ induced by the magnetic field, the emission timescale of the 
source $T_{\rm S}$, as compared to the lifetime of the experiment.
For each case, we simulated clusters of UHE CRs using the 
instrumental characteristics typical of future experiments such
as the Telescope Array, the High Resolution Fly's Eye, and most
notably, the Pierre Auger 
Project. To this end we have simulated the emission, the propagation in
an extra-galactic magnetic field, and the detection of clusters of UHE
CRs. We then performed a Monte-Carlo likelihood analysis on these
UHE CRs, and tried to reconstruct the physical parameters from the
maximum of the likelihood. We simulated clusters of $\sim40$ events, 
as the next generation experiments are expected to detect
$\sim20-100$ events per cluster if the clustering recently
suggested by the AGASA experiment~\cite{Hayashida1} is real.

  In summary, the likelihood presents different degeneracies between
different parameters, which complicates the analysis. As an example,
the likelihood is degenerate in the ratios $N_0/T_{\rm S}$, or
$N_0/\Delta\tau_{100}$, where $N_0$ is the total fluence, and
$\Delta\tau_{100}$ is the spread in arrival time: these ratios
represent rates of detection. Another example is given by the
degeneracy between the distance $D$ and the injection energy 
spectrum index $\gamma$. Yet another is the ratio
$(D\tau_E)^{1/2}/l_c$,
that controls the size of the scatter around the mean of the 
$\tau_E-E$ correlation. Therefore, in most general cases, values for 
the different parameters cannot be pinned down, and generally, only 
domains of validity are found.

  We find that the distance to the source is obtained from the pion 
production signature, above the Greisen-Zatsepin-Kuzmin cut-off, when 
the emission timescale of the source dominates over the time delay. 
Since the time delay decreases with increasing energy, we find
that the lower the 
energy $E_{\rm C}$, defined by $\tau_{E_{\rm C}}\simeq T_{\rm S}$, the 
higher the accuracy on the distance $D$. The error on $D$ is, in the 
best case, typically a factor 2, for one cluster of $\simeq40$ events.
In this case, where the emission timescale dominates over the time 
delay at all observable energies,
information on the magnetic field is only contained in the
angular image, which we did not systematically include into our
likelihood analysis due to computational limits. A more detailed
investigation of angular images will be presented separately in
a forthcoming study. Qualitatively, the size of the angular image is 
proportional to $B_{\rm rms}(Dl_c)^{1/2}$, whereas the structure
of the image, {\it i.e.} the number of separate images, is
controled by the ratio $D^{3/2}B_{\rm rms}/l_c^{1/2}$. Finally,
the case where the time delay 
dominates over the emission timescale, with a time delay shorter than 
the lifetime of the experiment, also allows to estimate the distance 
with a reasonable accuracy. 

  The strength of the magnetic field can only be obtained from the 
time-energy image in this latter case because the angular image
will not be resolvable. When the time delay 
dominates over the emission timescale, and is, at the same time,
larger than the lifetime of the experiment, only a lower limit 
corresponding to this latter timescale, can be placed on the time 
delay, hence on the strength of the magnetic field. When combined with 
the Faraday rotation upper limit,
this would nonetheless allow to bracket the r.m.s. magnetic
field strength within a
few orders of magnitude. Here as well, significant information is
contained in the angular image.

  The coherence length enters the ratio $(D\tau_E)^{1/2}/l_c$ that 
controls the scatter around the mean of the $\tau_E-E$ correlation in 
the time-energy image. It can therefore be estimated from the width of 
this image, provided the emission timescale is dominated by $\tau_E$ 
(otherwise the correlation would not be seen), and some prior 
information on $D$ and $\tau_E$ is available.

  As a concluding remark, we point out that the magnetic field, 
although it 'scrambles' the images of UHE CRs, also brings 
extra-information. It not only leaves a signature of its own, it may 
also, in the case where the time delay becomes comparable to the 
emission timescale at some intermediate energy, allow an evaluation of 
the emission timescale of the source itself. There is therefore very 
important information hiding in the angle-time-energy images of UHE 
CRs, which could be exploited by future large-scale experiments.

\section*{Acknowledgments}
Peter Biermann, Pasquale Blasi, Chris Hill, and
J\"org Rachen are acknowledged for valuable discussions. We
especially thank Jim Cronin for
encouraging us to perform this study and Angela Olinto and David
Schramm for collaboration in earlier work on which it is
partly based. We are grateful to the
Max-Planck Institut f\"ur Physik, M\"unchen, Germany for
providing CPU time. G.S. acknowledges financial support by the Deutsche
Forschungs Gemeinschaft under grant SFB 375 and by the
Max-Planck Institut f\"ur Physik. This work was supported, in
part, by the DoE, NSF, and NASA at the University of Chicago.

\newpage

\end{document}